# Developing a Method to Determine Electrical Conductivity in Meteoritic Materials with Applications to Induction Heating Theory

Daniella N. DellaGiustina
Physics 498: Senior Capstone Project Thesis

Advisor: Dr. Dante Lauretta
University of Arizona, Tucson
Spring 2008

Magnetic induction was first proposed as a planetary heating mechanism by Sonett and Colburn in 1968, in recent years this theory has lost favor as a plausible source of heating in the early solar system. However, new models of proto-planetary disk evolution suggest that magnetic fields play an important role in solar system formation. In particular, the magneto-hydrodynamic behavior of proto-planetary disks is believed to be responsible for the net outward flow of angular momentum in the solar system. It is important to re-evaluate the plausibility of magnetic induction based on the intense magnetic field environments described by the most recent models of proto-planetary disk evolution.

In order to re-evaluate electromagnetic induction theory the electrical conductivity of meteorites must be determined. To develop a technique capable of making these measurements, a time-varying magnetic field was generated to inductively heat metallic control samples. The thermal response of each sample, which depends on electrical conductivity, was monitored until a thermal steady state was achieved. The relationship between conductivity and thermal response can be exploited to estimate the electrical conductivity of unknown samples. After applying the technique to various metals it was recognized that this method is not capable of making precise electrical conductivity measurements. However, this method can constrain the product of the electrical conductivity and the square of the magnetic permeability, or $\sigma\mu^2$, for meteoritic and metallic samples alike. The results also illustrate that along with electrical conductivity $\sigma$, the magnetic permeability $\mu$ of a substance has an important effect on induction heating phenomena for paramagnetic ($\mu/\mu_0 > 1$) and especially ferromagnetic materials ($\mu/\mu_0 \gg 1$).

# 1. Introduction

Unlike typical semi-conductors, ordinary chondrites have 5-20 wt. % metal content in their composition (McSween et. al., 1999). The most significant contribution of electrical conductivity in these meteorites comes from the isotropic distribution of metals (considered "good conductors") within an otherwise semi-conducting compositional matrix. An appropriate technique to constrain the electrical conductivity of these materials will involve simultaneously measuring the conductivity of metallic and semi-conducting minerals present in a meteoritic sample. This presents a challenge since the conductivity of metals and semi-conductors, especially silicates, are often measured using distinctly different techniques. In order to constrain the electromagnetic properties of semi-conducting meteoritic materials a new technique must be developed.

The primary goal of the experiments discussed in this paper was to develop a technique to determine electrical conductivity of metallic control samples, which already have well defined electromagnetic properties. If this technique proved accurate for metallic samples, it could be applied to semi-conducting meteoritic samples which have unknown electrical conductivities.

# 2. Magnetic Induction of Planetismals

Determining the inherent properties of meteoritic materials has significant scientific value. Meteorites represent the most primitive matter in the solar system; these objects come from parent-body asteroids which are the remnants of terrestrial planet formation. Understanding the properties of such primitive objects can provide insights into the chemical and physical processes which occurred during solar system formation.

In particular, the petrography of meteorite samples indicates that most parent-body asteroids underwent thermal metamorphism during formation. Although this metamorphism varies from low-temperature aqueous alteration to complete melting and differentiation, it reveals the presence of an intense heat source during the formation of the solar system.

Magnetic induction of asteroids was first proposed as a heating mechanism by Sonett and Colburn in 1968, and extended upon by Herbert and Sonett in 1979. Induction heating occurs when an object is placed in a time varying magnetic field; the object will release thermal energy as it resists the induced current passing through it. The theory proposed a strong magnetic field anchored in the T-Tauri sun propagated outward through a high velocity, dense, and isotropic solar wind. As the solar wind and the embedded magnetic field came into contact with accreting planetismals, a network of electrical current was induced within the body. Heating would be concentrated along these current paths within the asteroid. Effectively, the kinetic energy associated with solar outflows is converted into thermal energy within the conducting asteroid via electromagnetic interaction.

Magnetic induction fell out of favor as a planetary heating mechanism in the early 1990s. During the same time, the role that magnetic fields play in solar system formation began to make serious advances within astrophysics. In particular, the concept of magneto-rotational instability was introduced as a mechanism of angular momentum transport.

The necessity of angular momentum transport is currently a challenge in modeling solar system formation. The most widely accepted theory of solar system formation, the nebular hypothesis, predicts that the solar system formed from the gravitational self-collapse of a dense cloud of gas and dust. After collapse, the proto-Sun formed by this event must increase its rotation rate due to conservation angular momentum. However, as the solar system formed angular momentum was transferred by some mechanism from the Sun outward; the present-day Sun only maintains one percent of the angular momentum in the solar system. One of the most widely accepted mechanisms for the net outward flow of angular momentum is that of magneto-rotational instability within a shearing, magneto-hydrodynamic disk.

Magneto-rotational instability (MRI) predicts that in the presence of a rotational sheer, magnetic fields can grow very rapidly within a Keplarian disk. If a disk becomes ionized, a weak magnetic field will arise between gas parcels in the disk. This magnetic field lines act elastically, linking gas parcels together. As the gas parcel closest to the proto-sun begins to orbit more quickly, a consequence of Kepler's laws, the magnetic field linking the parcels will stretch. This will cause the outer gas parcel to produce a retarding force on the forward motion of the inner parcel. This retarding force will cause some angular momentum to transfer from the inner to the outer gas parcel. As the inner parcel loses angular momentum, however, it will fall further towards the proto-sun with increased angular velocity, causing more stressing of the magnetic field lines and therefore increasing the magnetic force present. Because this process repeats itself with positive feedback, a strong magnetic field can grow exponentially from weakly ionized gases. (Balbus and Hawley, 1991).

This process has been modeled extensively; in particular Hawley et al. developed a three dimensional magneto-hydrodynamic (MHD) code to compute the evolution of MRI field strength. Their results showed that magnetic field produced by weakly ionized gases can rapidly grow into a turbulent disk. This turbulence is results in a net outward flow of angular momentum, and is maintain by magneto-rotational instability. As this process evolves magnetic field strength will reach a semi-steady state. Studies conducted to quantify mean strength predict that fields from 200 mG to 10 G exist within proto-planetary disks (Salmeron and Wardle, 2005). The upper limit of these field strengths has significant implications for planetary heating; such strong magnetic fields could easily induce electric currents, even in poorly conducting materials. The effect of such an environment on accreting planetismals has yet to be evaluated.

A prerequisite for MRI to arise within a disk is a local ionization source. Current observations of pre-main sequence young stellar objects (YSOs) reveal that magnetic reconnection flares (which exceed the strength of the strongest solar flares by orders of magnitude) are frequent and prevalent in these bodies. Recently, observations were made by the Chandra X-ray observatory of the COUP field in the Orion cluster; the field contains a suite of YSOs that are all comparable in mass to the sun.

Examining the X-ray emissions from the COUP field revealed that these objects emitted frequent, highly energetic stellar flares (Favata et al. 2005). These flares can reach ~0.5 AU in length, occurred approximately once every six days and lasted up to 3 days. Analyses conducted to quantify the minimum magnetic field strength necessary to confine the flaring plasma infer that these objects have magnetic field strengths of up to 3000 G. (by comparison, the Sun's average magnetic field is only 1 G). Additional studies have suggested that energetic baryonic particles produced by these X-ray flares could interact with disk material to produce short lived radionuclides, such as Al-26 (Chaussidon and Gounelle, 2006).

Induction heating of planetismals has yet to be investigated in the environment of MHD turbulent disks. Both accretion and MRI processes are predicted to be active during same era in proto-planetary disk evolution. Therefore, to appropriately evaluate the role of magnetic induction heating in the early solar system, this scenario must be considered. The environment described by MRI in MHD turbulent disks might actually prove more suitable for magnetic induction heating than the environment first described by Sonett when the theory was proposed. Further research must be conducted in order to evaluate how much heating magnetic induction could have contributed to the early solar system environment.

Although no longer considered a significant source of thermal metamorphism in planetismals, the plausibility of magnetic induction heating should be re-evaluated in light of the advancements made within astrophysics. Theoretical models suggest that magnetic fields play an important role in the evolution of proto-planetary disks. To properly re-evaluate this theory, experimental simulations must be conducted to investigate the behavior of chondritic material within a known magnetic field. This will require detailed numerical modeling to determine how much heat is produced as function of electrical conductivity.

Unfortunately, there has been little work to computationally simulate these processes because the electromagnetic properties, especially conductivity, of the material present during accretion are not well known. Determining the electrical conductivity of various meteorites is therefore a significant contribution to the field of planetary sciences.

## 3. Experimental Theory

Developing a technique which utilizes induction heating to determine electrical conductivity of materials is of particular interest. Samples used to determine the electrical conductivity could also be observed directly for any morphological or compositional changes that might occur during the induction process.

During induction heating, the electromagnetic and thermal properties are tightly coupled in a non-linear way. This relationship can be described by modifying Fourier's equation (Dughiero et. al, 1996), in cylindrical coordinates, to our experimental assumptions:

$$\kappa \nabla T(r, z, \phi) - C_p \rho \frac{\partial T}{\partial t} = \frac{\sigma \omega^2}{c^2} A^2 \qquad (1)$$

Where $\kappa$ is the thermal conductivity of the sample, T the sample temperature, $C_p$ the specific heat, $\rho$ the mass density, $\sigma$ the electrical conductivity, $\omega$ the applied frequency, $c$ the speed of light, and $A$ the amplitude of the magnetic vector potential from the applied time-varying magnetic field. Additionally, the solution must meet the following boundary condition:

$$\kappa \frac{\partial T}{\partial r} = \varepsilon \sigma_s (T_s^4 - T_a^4) \qquad (2)$$

In this expression, $\frac{\partial T}{\partial r}$ denotes the temperature gradient normal to the surface. Where $\varepsilon$ is the emissivity of the sample, $\sigma_s$ the Stefan-Boltzman constant, $T_s$ the surface temperature at any given point on the sample, and $T_a$ is the ambient temperature of the sample.

Knowing the spatial and temporal dependence of the applied magnetic field, and monitoring the temperature of the sample as a function of time, it is possible to determine the constants in equation (1) using numerical Monte Carlo methods (Dughiero et. al., 1996). Using numerical methods is not always useful to constrain these constants since less complicated approaches have been developed to determine these parameters. However for electrical conductivity, this method is able to simultaneously measure contributions from both metallic and semi-conducting minerals in meteorites. This method is therefore well-suited to determine the bulk electrical conductivity of both ordinary chondrites and pure metals.

Unfortunately, the equipment necessary to monitor the temperature of inductively heated samples is costly and not readily available. To apply these numerical methods, it is necessary to map the temperature distribution along the surface of a sample in sub-millimeter resolution. Therefore a less sophisticated approach was taken to obtain electrical conductivity of both ordinary chondrites and pure metals by making the following hypothesis: Electrical conductivity will have more effect on the thermal response of an inductively heated sample than any other intrinsic property of the material.

Referencing equation (1), the following assumptions were made, and the hypothesis was expanded upon: If an inductively heated sample is allowed to reach thermal steady state, the affects of inherent properties such as thermal conductivity $\kappa$, specific heat $C_p$, and mass density $\rho$ will no longer influence the average steady state temperature. Additionally, if several samples with the same geometry are placed in the same time varying magnetic field (the frequency $\omega$, and vector potential $A$, remain consistent) then the difference in their thermal response will only depend on differing electrical conductivity. Measuring the thermal response of several samples of known conductivities under these conditions will illustrate the relationship between average steady state temperature and conductivity. Experiments were performed to establish this relationship for pure metals, which have well defined electromagnetic properties. Figure 1 represents a *predicted* relationship between the steady state temperature and electrical conductivity of several metallic samples.

Once the mathematical relationship between thermal response and electrical conductivity is established, the conductivity of unknown samples (both metallic and semi-conducting) can be interpolated using the same relationship. Unknown samples will be inductively heated under identical conditions as the control samples, and their thermal response will be recorded and used to estimate electrical conductivity, σ.

## 4. Experimental Details

Experimental samples (pure metals of cylindrical geometry) were placed in a silica sample chamber which was evacuated prior to each experiment. Running the experiments under vacuum is necessary to avoid both sample oxidation and heat transfer via convection. The evacuated sample chamber was then suspended in an Ameritherm HotShot® radio frequency (RF) induction heating station. This device is designed to heat objects by generating frequencies from 150 to 400 kHz with up to 2 kW of power. Using AC currents (up to 300 A) the magnetic field is produced using various water-cooled copper induction heating coils. The field strength produced by the primary coil, designed specifically for these experiments, can be approximated by the mathematical expression for an infinite solenoid

$$\vec{B} = \mu_0 n I \quad (3)$$

Where $\mu_{air} \approx \mu_0$ is the permeability of free space, n is the number of turns per unit length, and I is the current through the coil. The AC current running through the coil can be expressed as:

$$I = I_0 \sin(\omega t) \quad (4)$$

Where $I_0$ is the maximum current, which can be programmed using the RF furnace control panel, and ω is the angular frequency (given by the control panel). Using this experimental setup, the primary coil is capable of generating magnetic fields up to 990 Gauss.

Sample temperature was monitored using an Omega iR2 pyrometer and an Omega OS37-10-K pyrometer. Optical pyrometers, or infrared thermometers, are non-contact temperature sensors that are will suited for applications involving metals and semiconductors. The iR2 pyrometer is designed for high temperature (450-3000 ºC) applications. It has an accuracy of 0.2% of full scale when placed no more than 4 m from the sample. Temperature is measured using a 2-color ratio technique in which is essential for accurate measurements when the emissivity of the target is unknown, or when the target is viewed through another medium which reduces energy. The OS37-10-K is used to measure temperatures below 450 °C. It has a response time of 1 sec and an accuracy of 2% when placed no more than 10 cm from the sample. An internal laser can be used for targeting to ensure that sample temperature is being monitored.

A side view diagram (Figure 2) and photographs of the experimental setup (Figures 3 & 4) are included in the appendix.

During an experimental run, the metallic sample was placed in the silica chamber, and then placed in the heating station coil. The chamber was evacuated, and the pyrometer positioned to monitor sample temperature. After staring the water cooling system, the RF induction furnace was programmed to the desired current and allowed to run. Once the sample appeared to reach thermal steady state the average temperature was recorded.

Two metals of differing conductivities were available for the experiments: silver (Ag) and iron (Fe). Only cylindrical samples of high purity Ag and Fe were run in the experiments. These samples were approximately 0.25 inches in diameter and 0.50 inches in length. For future applications, meteoritic samples can be cut to the same dimensions using specific coring drill bits.

## 5. Experimental Results

Table 1 represents the results obtained during the first experimental run, where $T_s$ denotes the average steady state temperature. The uncertainty in $T_s$ is from the fluctuations in the steady state temperature measurements, not the uncertainty associated with the optical pyrometers which is significantly less. Values of metallic conductivity were obtained from Serway, 1998.

| Metallic Sample | σ (M S/m) | $I_0$ (A) | f (kHz) | $T_s$ (C°) | |
|---|---|---|---|---|---|
| Ag | 62.9 | 277 A | 200-305 | 90 | ±10 |
| Fe | 10 | 277 A | 215-320 | 1200 | ±50 |

*Table 1: Properties of Experimental Samples*

The high temperatures achieved while running Fe caused the sample chamber to fuse to the Fe sample. As a result, the sample chamber cracked as it cooled to room temperature, preventing additional experimental runs.

## 6. Discussion

The experimental results indicate that electrical conductivity is *not* the most significant intrinsic property when a sample is inductively heated. To determine the why the hypothesis posed was incorrect, it is necessary to go back to the first principles of electromagnetism.

To calculate the thermal energy delivered to each sample from magnetic induction, the power (energy per unit time) should be determined. To solve for the power generated in a sample, the electric field, $\vec{E}$, produced by the induced current *in the sample* must be first be determined. This can be found using Faraday's Law of Induction:

$$\nabla \times \vec{E} = -\frac{\partial \vec{B}}{\partial t}$$

Rewritten in integral form as:

$$\oint \vec{E} \cdot d\vec{s} = -\frac{\partial \Phi_B}{\partial t} \quad (5)$$

Where $\Phi_B$ is the magnetic flux through the cylindrical sample. For a sample of permeability $\mu$, and an applied magnetic field described by equations (3) and (4), the magnetic flux *through* the sample can be expressed as a function of $r$:

$$\Phi_B = \pi r^2 \mu n I_0 \sin(\omega t) \quad (6)$$

Inserting this expression into equation (5) will yield the following result:

$$2\pi r \vec{E} = -\frac{\partial}{\partial t}\left(\pi r^2 \mu n I_0 \sin(\omega t)\right)$$

This is rearranged to express the electric field produced by the current induced in the sample:

$$\vec{E}_{induc} = -\frac{\mu n \omega r}{2} I_o \cos(\omega t) \quad (7)$$

The induced power can be found using Ohm's Law and performing cosmetic algebra:

$$\frac{P}{V} = I^2 R = \frac{J^2}{\sigma} = E^2 \sigma$$

For cylindrical samples, this is expressed as the following integral:

$$P = \int_V dV\, E^2 \sigma = \int_0^H \int_0^R E^2 \sigma\, 2\pi r\, dr\, dz$$

Where $H$ is the height of the sample and $R$ is the radius. Inserting equation (7) into this expression and solving the integral results in the following:

$$P = \frac{\pi}{8} H R^4 \sigma \left(\mu n \omega I_o \cos(\omega t)\right)^2 \quad (8)$$

The time-average of $\cos^2(\omega t) = \frac{1}{2}$, therefore the average power delivered to the sample via electromagnetic induction is found to be:

$$P_{avg} = \frac{\pi}{16} H R^4 \sigma \left(\mu n \omega I_o\right)^2 \quad (9)$$

Equation (9) illustrates that conductivity σ, is *not* the most relevant intrinsic material property in induction heating applications for materials with relative magnetic permeabilities that deviate from unity. Although the average power is related to linearly to σ, it is also proportional to the square of the sample's magnetic permeability μ. Therefore, if a suite of samples with the same geometry ($H$ and $R$ remain constant) are placed in the same magnetic field ($n$, $I_o$, and the range of ω are consistent), the thermal response of the sample will depend on the term σμ².

Magnetic permeability, μ, is related to the magnetic susceptibility of a material $\chi_M$, by the following relationship:

$$\mu = \mu_0 (1 + \chi_M)$$

Although the electrical conductivity of Ag is higher than any other metal, it is diamagnetic and therefore has a small, negative magnetic susceptibility represented by a scalar quantity. Likewise Fe is ferromagnetic, having a large, positive magnetic susceptibility represented by a differential tensor (Polluck and Stump, 2002). The experimental results are therefore completely consistent with the *correct* theoretical assumptions.

## 7. Conclusion

Although the technique developed is not well suited to experimentally determine the relationship between thermal response $T_s$, and conductivity σ, in induction heating applications, it can determine the relationship between $T_s$ and σμ². Obtaining the product of two intrinsic material properties is not as desirable as obtaining these values separately, however this method is still useful. Because there is virtually no data regarding the electromagnetic properties of meteorites, determining the parameter σμ² for ordinary chondrites is still of scientific value.